# Toward Economics as a New Complex System

Taisei Kaizoji

Graduate School of Arts and Sciences,

International Christian University

Abstract

The 2015 Nobel Prize in Economic Sciences was awarded to Eugene Fama, Lars Peter Hansen and Robert Shiller for their contributions to the empirical analysis of asset prices. Eugene Fama [1] is an advocate of the efficient market hypothesis. The efficient market hypothesis assumes that asset price is determined by using all available information and only reacts to new information not incorporated into the fundamentals. Thus, the movement of stock prices is unpredictable. Robert Shiller [2] has been studying the existence of irrational bubbles, which are defined as the long term deviations of asset price from the fundamentals. This drives us to the unsettled question of how the market actually works.

In this paper, I look back at the development of economics and consider the direction in which we should move in order to truly understand the workings of an economic society.

## Establishment of economics as a mathematical science

General equilibrium theory, originated by Leon Walras [3], is a starting point for the development of economics as a mathematical science. It indicates that aggregate demand accords simultaneously with aggregate supply in many competitive markets. Walras proposed an auction process (tatonnement process) formalizing Adam Smith's theory of the "invisible hand", according to which unobservable forces adjust the equilibrium price to match supply and demand in a competitive market.

Paul Samuelson took a hint for his economic theories from classical mechanics, as first presented by Issac Newton (Lagrangian analytical mechanics) [4]. More concretely, by applying the method of Lagrange multipliers, Samuelson mathematically formulated the behavior of economic agents as a constrained optimization problem in his *Foundations of Economic Analysis* published in 1947. He also mathematically formulated Walras' tatonnement process. Economics became a mathematical science largely through his work and insights.

Thereafter, the first rigorous mathematical proofs of the existence of Walrasian equilibrium and the stability of that equilibrium were given by Gérard Debreu [5], K. Arrow [6] and others. Thus, Walrasian general equilibrium theory was completed through the efforts of many notable economists, forming the foundation for the development of neo-classical economics.

**Rational Expectations and DSGE model**

In the 1960s, John F. Muth [7] advanced the theory of rational expectation. The rational expectation hypothesis is defined as an all available information hypothesis for human anticipation, according to which an economic agent's anticipation (expectation) is the best predictor of future conditions. The first scholars to pay much attention to the idea of rational expectation were Robert Lucas [8] and Thomas Sergent [9]. Lucas introduced Muth's line of thinking into macroeconomics and aggressively promoted the rational expectation revolution among his colleagues. Although Muth originally believed that rational expectation holds only in isolated markets, following the so-called Lukas critique [10], many macroeconomists concluded that economic agents who obtain all available information are able to recognize the general equilibrium in the correct macroeconomic model. It is worth noting that following the completion of Walrasian general equilibrium theory, dynamic general equilibrium models were formulated as dynamic optimization problems for economic agents by Frank P. Ramsey [11]. The Ramsey model is an application of Hamilton's vibrational principle in classical mechanics in macroeconomics. Assuming that an individual decides his/her current behavior by following a predicted future course, it seems that the introduction of static general equilibrium theory into dynamic general equilibrium theory is a necessary consequence.

One problem with dynamic general equilibrium models is that the intertemporal equilibrium is a saddle-point type of equilibrium. The only way to reach the intertemporal equilibrium is to select the initial value on one of the stable branches of the saddle point. Otherwise, this results in a failure to attain equilibrium. The question, then, is how to choose an initial value that will lead to the sought-after equilibrium? Rational expectation is the key to answering this question. If economic agents— consumers and firms—have rational expectations, that is, if these agents are familiar with the dynamic general equilibrium model and have a forecast of the equilibrium coinciding, on average, with the intertemporal equilibrium, then the rational expectation equilibrium will be basically equivalent to the intertemporal equilibrium. The theory of pre-established harmony in economics was completed in this way.

Subsequently, real business cycles theory [12] and new Keynesian macroeconomics [13] have developed as classes of the dynamic stochastic general equilibrium (DSGE) model, and the DSGE model has become dominant in macroeconomics.

**Failures of the DSGE models**

However, it is clear that dynamic general equilibrium models that include the rational expectation formulation have problems explaining many 'real world' economic phenomena. In this section, we examine questions regarding these DSGE models.

To begin with, the Ramsey model and a number of DSGE models assume a 'Robinson Crusoe economy'. In a Robinson Crusoe economy, there is a single representative agent (that is, a Robinson Crusoe). To maximize utility over his life as a consumer, this lone agent produces goods by using capital as a producer. In such an economy, there is no broader society or community and, therefore, no interaction with others.

These models for a planned economy controlled by a dictator rather than for a market-driven capitalist economy. The general public likely believes that macroeconomics studies macro phenomena generated by complicated interactions between microelements—phenomena that might include changes in consumer behavior and company activity—but this appears to be the most popular of all delusions regarding advanced macroeconomics.

Furthermore, DSGE models are faced with two technical difficulties. First, they assume that the economic agent is able to predict the general equilibrium correctly and that the economy will follow along a path toward that equilibrium. However, the rational expectation equilibrium cannot be solved for analytically in a number of DSGE models. Since DSGE models are nonlinear, expressed in nonlinear functions such as utility functions and production functions, multiple equilibria are possible. Thus we confront the problem of which equilibrium should be chosen by the economic agent. In short, even the authors of DSGE models are not able to tell precisely the direction in which the economy will proceed.

In addition, nonlinear DSGE models are linearized around the equilibrium in the final stage. Generally the linearized model is effective only in the case of small shocks; once a large shock occurs, the linearized DSGE models are likely to be powerless to explain what is happening in the real economy. Given this limitation, it is no wonder that the DSGE models were unable to predict the impact on the macro economy of the global financial crisis of 2008 [14].

DSGE models omit, or treat lightly, the heterogeneity of economic agents and the interaction between and among these agents. The mistake that the theory makes is in assuming that the mechanism of economic phenomena at the macro-level can be replaced with rational behavior as seen from a micro-viewpoint. It seems that 'real' people live in an economy far from the economy supposed by advanced macroeconomics. As researchers who have objected to DSGE models have pointed out, macroeconomics needs to be transformed from a study of the economy of a representative agent's microscopic world into a study of macroscopic phenomena created by an influencing interaction between heterogeneous economic agents. (For example, see [15].)

**The new tide of macroeconomics: Microeconomic origin of macroeconomic tail risks**

The new tide of macroeconomics is based on research focused on the large macroeconomic fluctuations that originate from microeconomic shocks, as proposed by [16] and [17].Gabaix [16] demonstrates that microeconomic shocks to large firms affect aggregate outcomes when the distribution of firm sizes is very fat-tailed. Acemoglu et al. [17] demonstrates that when the empirical distribution of degrees of the intersectoral network corresponding to the U.S. input–output matrix can be approximated by a power law distribution, then microeconomic shocks to firms or disaggregated sectors can generate large aggregate fluctuations. These are macroeconomic phenomena that DSGE models overlook. While the work of these researchers assumes Zipf's law for company size and a heavy-tailed distribution of degrees of the intersectoral network, it does not answer the question of why the disparities occur.

**Econophysics: What we should do to truly understand truly macroeconomic phenomena**

DSGE models succeed in simplifying human behavior by using the assumption of rational expectation. However, a human being is essentially a social animal, and the essence of an economic society is the interaction between and among economic agents. In a macro-economy, important phenomena are generated by such interactions. Econophysics examines these macroscopic phenomena, including, for example, power laws for income, wealth, company size, stock prices, land prices, financial indicators, asset returns, and degrees of various economic and social networks. In addition, Econophysics studies phenomena caused by the collective psychology of economic agents such as speculative bubbles and subsequent crashes in asset markets. (For more detail regarding Econophysics, see, for example, introductory books [18-22]).

Econophysics attempts to elucidate the mechanism of the so-called "winner-take-all" or "rich get richer" phenomenon by applying statistical mechanics. In my view, Econophysics points to an alternative way of considering problems in the economic sciences. However, it seems that Econophysics is ignored by the great majority of economists even though a number of articles have been published in prestigious academic journals such as *Nature* and Physical Review Letters.

What is the barrier to a wider diffusion of Econophysics? I believe that statistical mechanics is an indispensable tool to understand the collective behavior generated by interactions between economic agents. However, merely replacing the behavior of an economic agent with the spin of an atom is insufficient as economic science. It appears to me that the reason why the impressive work of those engaged in Econophysics is not seriously discussed by the economist community is that the basic principles of Econophysics have not been fully demonstrated. For broader acceptance, it will be necessary for researchers to present their commentary on the theories of Econophysics using the terminology of economics, as Samuelson did in his *Foundations of Economic Analysis.*

### A principle of Econophysics

A number of Econophysics researchers model the behavior of an economic agent by using spin models such as the Ising model. These spin models are modeled by the principle of maximum entropy, and the configuration probability of the Ising model is given by the Boltzmann distribution with inverse temperature. In an economics context, the Ising spin model is equivalent to the discrete choice model [23] formulated as the random utility maximization problem and extended into a model with interaction between economic agents by Brock and Durlauf [24]. As is well known, the optimal choice probability of the discrete choice models is identical formally to the configuration probability of the Ising model. In fact, that random utility maximization problem is identical mathematically to the entropy maximization problem can be demonstrated mathematically by applying conjugate theory (See Miyagi [26]). The discrete choice model assumes that the random utility of the economic agent follows a Gumbel distribution. One can also demonstrate that the parameter that determines the variance of the Gumbel distribution corresponds to thermodynamic temperature [27][1].

As described above, it is possible to establish a principle of Econophysics that translates the principle of maximum entropy into economics; this is what Samuelson omitted when he introduced physical methods into economics.

**Toward economics as a new complex system: Can economics be a physical science?**

We are now able to see that economic systems are complex systems composed of many interacting agents. The fact that, in economic and financial systems, power laws, which are generally considered to be a common property characterizing complex systems, are often found empirically suggests that our economic society is an unknown complex system. Of course, having suggested this, we still have a long way to go before we arrive at an understanding of the true mechanism generating economic and financial phenomena.

In conclusion, economics has infinite potential for development as a new field of complex systems beyond merely serving as an opportunity to apply statistical mechanics.

---

[1] In addition, one may formulate the random utility maximization problem using the Tsallis entropy [25] that is a generalization of the Boltzmann–Gibbs entropy.